\documentclass[conference]{IEEEtran}
\usepackage{color}
\usepackage{pifont}
\usepackage{bm}
\usepackage{amsfonts}
\usepackage{amssymb}
\usepackage{flushend }
\usepackage{graphicx}
\usepackage{dsfont}
\usepackage{subfigure}
\usepackage{subeqnarray}
\usepackage[cmex10]{amsmath}
\usepackage{algorithmic}
\usepackage[ruled,linesnumbered]{algorithm2e}
\usepackage{txfonts}
\usepackage{multirow}

\begin{document}

\title{The APC Algorithm of Solving Large-Scale \\ Linear Systems: A Generalized Analysis \thanks{This work is sponsored in part by the National Natural Science Foundation of China (grant no. 61971058, 61801048, 61631004, 62071063) and Beijing Natural Science Foundation (grant no. L202014, L192002).}}

\author{\IEEEauthorblockN{Jiyan~Zhang, Yue~Xue, Yuan~Qi, and Jiale~Wang}
\IEEEauthorblockA{\textit{Beijing Univ. of Posts and Telecommunications (BUPT)}\\
Beijing, China \\
e-mail:\{zhangjiyan, runfordream, qiyuan, wangjiale\}@bupt.edu.cn}}

\maketitle

\begin{abstract}
A new algorithm called accelerated projection-based consensus (APC) has recently emerged as a promising approach to solve large-scale systems of linear equations in a distributed fashion. The algorithm adopts the federated architecture, and attracts increasing research interest; however, it's performance analysis is still incomplete, e.g., the error performance under noisy condition has not yet been investigated. In this paper, we focus on providing a generalized analysis by the use of the linear system theory, such that the error performance of the APC algorithm for solving linear systems in presence of additive noise can be clarified. We specifically provide a closed-form expression of the error of solution attained by the APC algorithm. Numerical results demonstrate the error performance of the APC algorithm, validating the presented analysis.
\end{abstract}

\begin{IEEEkeywords}
Large-scale systems, linear equations, distributed algorithms, performance analysis.

\end{IEEEkeywords}

\IEEEpeerreviewmaketitle

\section{Introduction}

\IEEEPARstart{S}olving large-scale systems of linear equations is a fundamental problem in various signal processing, control theory, and machine learning applications \cite{Azizan_2019}-\cite{Zuo_2021}. Recently, \emph{Azizan-Ruhi}, \emph{Avestimehrm}, and \emph{Hassibi} developed the APC algorithm for distributed solution of large-scale systems of linear equations \cite{Azizan_2019}. APC offers significant speed-up relative to other distributed methods such as the well-known alternating direction method of multipliers (ADMM) method \cite{Azizan_2019}, making the algorithm appealing for many applications\cite{Azizan_2019,Zuo_2021,Jia_2021}. The APC algorithm was used by the probabilistic load flow calculation of power systems for the privacy-preserving purpose \cite{Jia_2021}. Inspired by the APC algorithm, our previous work designed a distributed channel estimation algorithm for mmWave massive MIMO communication systems \cite{Zuo_2021}. Moreover, there are a number of references on the topics related to the APC algorithm, e.g., distributed algorithms for systems of linear equations \cite{Alaviani_2021}, for state estimation \cite{Zivojevic_2021}, for gradient-descent method \cite{Chakrabarti_2020,Chakrabarti_2021}, for linear transforms \cite{Dutta_2016,Dutta_2019}, for coded matrix multiplication \cite{Dutta_2020}, and for phase retrieval \cite{Zhao_2018}.

One distinguishing attribute of the APC algorithm is that it adopts the \emph{federated} architecture (also known as the \emph{server-based} distributed architecture \cite{Federated_Li_2020,Chakrabarti_2020}). The federated architecture comprises of one server (master) and multiple agents (workers) \cite{Chakrabarti_2020}-\cite{Dutta_2020}. Research attention has increasingly focused on the development of novel algorithms to distributed computation \cite{Alaviani_2021}\cite{Chakrabarti_2020}-\cite{Zhao_2018}, including those with the federated architecture \cite{Chakrabarti_2020}-\cite{Dutta_2020}; due to this reason, the APC algorithm attracts increasing research interest \cite{Zivojevic_2021,Chakrabarti_2020}-\cite{Zhao_2018}. 

However, the seminal work on the APC algorithm \cite{Azizan_2019} only considers a model setting wherein the system is free from noise, while noise is an unavoidable factor in real systems. It is still unconfirmed whether the APC algorithm can effectively solve linear systems with the existence of noise; in other words, researchers still lack a clear understanding of the robustness of the algorithm under the noisy condition. Therefore, this study aims at deriving the analytical results by considering the additive noise. Specifically, we generalize the analysis of the APC algorithm by utilizing the linear system theory (see Theorem 1), such that the error performance of the algorithm for solving linear systems is clarified in presence of noise (see Theorem 3). Note that the study in this paper derives analytical results which can be reduced to those in \cite{Azizan_2019}, by letting the elements of noise vector be zeros; in this sense, the analysis conducted in this paper is the generalized one. Moreover, we provide closed-form expressions to two important parameters of the APC algorithm (see Remark 1).

\emph{Notations}: Let $\mathbb{C}$ be the complex field. We write $\| \cdot \|_2$ for the $\ell_2$ norm of a vector. For a matrix $\textbf{A}$, $\lambda_{min}(\textbf{A})$ and $\lambda_{max}(\textbf{A})$ are the smallest and largest eigenvalues, respectively, and $\rho(\textbf{A})$ is the spectral radius of $\textbf{A}$, i.e., the largest absolute value of its eigenvalues. When $\textbf{A}$ has full column rank, we let $\textbf{A}^\dag = \left(\textbf{A}^H \textbf{A} \right)^{-1} \textbf{A}^H$ be the Moore-Penrose pseudoinverse of $\textbf{A}$. If $\textmd{span}(\textbf{A})$ is the span of columns in $\textbf{A}$, then $\textbf{P}_{\textbf{A}}^\bot = \textbf{I} - \textbf{A}\textbf{A}^\dag$ is the projection onto the orthogonal complement of $\textmd{span}(\textbf{A})$. For a block diagonal matrix $\textbf{A} = \textrm{diag}(\textbf{A}_{11}, \cdots, \textbf{A}_{kk})$, one can write $\textbf{A} = \textbf{A}_{11} \bigoplus \textbf{A}_{22} \bigoplus \cdots \bigoplus \textbf{A}_{kk}$ that is the direct sum of the matrices $\textbf{A}_{11}, \cdots, \textbf{A}_{kk}$ \cite{Horn_1985}. Let $\textbf{I}_N \in \mathbb{C}^{N \times N}$ be the identity matrix and $\textbf{0}_{N} \in \mathbb{C}^{N}$ ($\textbf{O}_{N} \in \mathbb{C}^{N \times N}$) be the all-zero vector (matrix).

\section{Model}

Consider the problem of solving a large-scale system of linear equations
\begin{eqnarray}\label{appx_L1_1}
  \textbf{A} \textbf{x} = \textbf{y},
\end{eqnarray}
where $\textbf{A} \in \mathbb{C}^{M \times s}$ has full column rank, $\textbf{x} \in \mathbb{C}^{s}$, and \begin{eqnarray} \label{appx_L1_2} \textbf{y} = \textbf{A} \textbf{x}^* + \tilde{\textbf{w}} \in \mathbb{C}^{M}.\end{eqnarray}
As usual, $\textbf{A}$ and $\textbf{y}$ are known matrix and vector, respectively. While $\tilde{\textbf{w}} = [\tilde{w}_1, \cdots, \tilde{w}_M]^T$ is an unknown noise vector (not necessarily Gaussian noise) in this paper.

We can apply the APC algorithm (presented in Algorithm \ref{algorithm2}) to find the solution to (\ref{appx_L1_1}), which will obtain an estimate of $\textbf{x}^*$. The formal description of APC \cite{Azizan_2019} is provided as Algorithm \ref{algorithm2}. The implementation of the APC algorithm requires $M$ distributed agents and one server. Every agent computes a solution to its own private equation, while all these agents can run in a parallel fashion. The server calculates the average of these $M$ solutions and regards it as the global solution. Initially, the APC algorithm executes the following computations: $\textbf{x}_\ell(0) =  \textbf{A}_{\ell}^H \left(\textbf{A}_{\ell} \textbf{A}_{\ell}^H\right)^{-1} y_\ell$ in every agent, and $\overline{\textbf{x}}(0) = \frac{1}{M} \sum_{\ell=1}^M \textbf{x}_\ell(0)$ in the server. After that, the computations are performed iteratively; that is, for $t=0, \cdots, T-1$, the APC algorithm goes through the following steps: $\textbf{x}_\ell(t+1) = \textbf{x}_\ell(t) + \gamma \textbf{P}_\ell^\bot \left(\overline{\textbf{x}}(t) - \textbf{x}_\ell(t) \right)$ and $\overline{\textbf{x}}(t+1) = \frac{\eta}{M} \sum_{\ell=1}^M \textbf{x}_\ell(t+1) + (1 - \eta ) \overline{\textbf{x}}(t)$, with the agents and server, respectively.

\begin{algorithm}[!h]
        \caption{\small The APC Algorithm for Finding the Solution $\tilde{\textbf{x}}$ to $\textbf{A} \textbf{x} = \textbf{y}$} \label{algorithm2}
            \KwIn{$\textbf{y} = \left[y_{1}, \cdots, y_{M}\right]^T \in \mathbb{C}^{M}$, $\textbf{A} \in \mathbb{C}^{M \times s}$, and $T$.}
            \KwOut{$\tilde{\textbf{z}} \in \mathbb{C}^{s}$.}
            Let $\textbf{A}_{\ell} \in \mathbb{C}^{1 \times s}$ denote the $\ell$-th row of $\textbf{A}$\;
            $t=0$\;
            \tcc{Computations in every agent:}
            \For{$\ell=1:M$}{
                $\textbf{P}_\ell^\bot = \textbf{I}_{s} - \textbf{A}_{\ell}^H \left(\textbf{A}_{\ell} \textbf{A}_{\ell}^H\right)^{-1} \textbf{A}_{\ell}$\;
                $\textbf{z}_\ell(0) =  \textbf{A}_{\ell}^H \left(\textbf{A}_{\ell} \textbf{A}_{\ell}^H\right)^{-1} y_\ell$\tcp*{Initializing each $\textbf{z}_\ell(0)$}
            }
            \tcc{Computations in the server:}
            $\overline{\textbf{x}}(0) = \frac{1}{M} \sum_{\ell=1}^M \textbf{x}_\ell(0)$\;
            \While{$t \leq T-1$}{
                \tcc{Computations in every agent:}
                \For{$\ell=1:M$}{
                $\textbf{x}_\ell(t+1) = \textbf{x}_\ell(t) + \gamma \textbf{P}_\ell^\bot \left(\overline{\textbf{x}}(t) - \textbf{x}_\ell(t) \right)$\;
                }
                \tcc{Computations in the server:}
                $\overline{\textbf{x}}(t+1) = \frac{\eta}{M} \sum_{\ell=1}^M \textbf{x}_\ell(t+1) + (1 - \eta ) \overline{\textbf{x}}(t)$\;
                $t=t+1$\;
            }
            $\tilde{\textbf{x}} = \overline{\textbf{x}}(t+1)$\;
\end{algorithm}

To clarify how to set the parameters $\gamma$ and $\eta$, we define \begin{eqnarray} \label{def_X_0425} \textbf{X} = \frac{1}{M} \sum_{\ell =1}^M  \textbf{A}_{\ell}^H \left(\textbf{A}_{\ell} \textbf{A}_{\ell}^H\right)^{-1} \textbf{A}_{\ell},\end{eqnarray} and denote the eigenvalues of $\textbf{X}$ by $\theta_i$, $1 \leq i \leq s$, where \begin{eqnarray} \label{def_theta_0425} \theta_s \leq \cdots  \leq \theta_1 \end{eqnarray} with $\theta_s = \theta_{min} = \lambda_{min}\left(\textbf{X} \right) \geq 0$ and $\theta_1 = \theta_{max} = \lambda_{max}\left(\textbf{X} \right) \leq 1$ \cite{Azizan_2019}. It is known from \cite[Theorem 2.5.6]{Horn_1985} that $\theta_1, \cdots, \theta_s$ are real-valued.

\emph{Remark 1}: The parameters $\gamma$ and $\eta$ in the APC algorithm are set as follows:
\begin{eqnarray} \label{eta_equ} &&
\gamma = \frac{2 \left(\sqrt{\theta_{max}} \sqrt{\theta_{min}} + 1 \right) - 2\sqrt{(1-\theta_{max})(1 - \theta_{min})}}{\left(\sqrt{\theta_{max}} + \sqrt{\theta_{min}} \right)^2}, \\
&&\eta = \frac{2 \left(\sqrt{\theta_{max}} \sqrt{\theta_{min}} + 1 \right) + 2\sqrt{(1-\theta_{max})(1 - \theta_{min})}}{\left(\sqrt{\theta_{max}} + \sqrt{\theta_{min}} \right)^2},
\label{gamma_equ}  \quad\quad
\end{eqnarray}
which satisfy \cite{Azizan_2019}
\begin{eqnarray} \label{eta_gamma_cond_equ}
&&\theta_{max} \eta \gamma =  \left(1 + \sqrt{(\gamma-1)(\eta-1)}\right)^2, \\
&&\theta_{min} \eta \gamma =  \left(1 - \sqrt{(\gamma-1)(\eta-1)}\right)^2.
\end{eqnarray}

Note here that the seminal work \cite{Azizan_2019} of the APC algorithm provided an indirect way of finding the optimal $\gamma$ and $\eta$, which requires to solve an optimizing problem and might only achieve near-optimal values in practical use. While in this paper we present closed-form expressions of the optimal $\gamma$ and $\eta$, i.e., (\ref{eta_equ}) and (\ref{gamma_equ}), respectively, so as to simplify the parameter setting as well as the forthcoming performance analysis.

Moreover, we define \begin{eqnarray} \label{alpha_def1} \alpha := \frac{\sqrt{\kappa(\textbf{X})}-1}{\sqrt{\kappa(\textbf{X})}+1},\end{eqnarray}where $\kappa \left(\textbf{X} \right) = \frac{\theta_{max}}{\theta_{min}} \geq 1$ that is the condition number of $\textbf{X}$ \cite{Azizan_2019}.

\section{Analysis}

In this section, we conduct a performance analysis of the APC algorithm, in terms of the error of solution (Definition 1). We are now going to present the main results whose proofs are given in Appendix.

\emph{Lemma 1}: Consider Algorithm \ref{algorithm2}, and assume that $\textbf{A}_\ell$ is the $\ell$-th row of $\textbf{A}$ (see Line 1). Then, the projection matrix onto the nullspace of $\textbf{A}_\ell^H$ can be expressed as
\begin{eqnarray}\textbf{P}_\ell^\bot = \textbf{I}_s - \textbf{A}_\ell^H \left( \textbf{A}_\ell \textbf{A}_\ell^H \right)^{-1} \textbf{A}_\ell, \end{eqnarray} which is calculated and used in Algorithm \ref{algorithm2} (see Line 4 and Line 10, respectively). Moreover, $\textbf{A}_\ell \textbf{P}_\ell^\bot = \textbf{0}_{L}^T$ and $\left(\textbf{P}_\ell^\bot\right)^2 = \textbf{P}_\ell^\bot$.

\emph{Definition 1}: Let \begin{eqnarray} \label{def_e_appx} &&\textbf{e}_\ell(t)= \textbf{x}_{\ell}(t) - \textbf{x}^*,  \\ &&\bar{\textbf{e}}(t)= \bar{\textbf{x}}(t) - \textbf{x}^* = \frac{1}{M} \sum_{\ell=1}^M \textbf{e}_\ell(t). \label{e_bar} \end{eqnarray} The key recursions in the APC algorithm, i.e., Line 10 and Line 12 of Algorithm \ref{algorithm2}, can be reformulated as
\begin{eqnarray} \label{e_recur1}
 && \textbf{e}_\ell(t+1) = \textbf{e}_\ell(t) + \gamma \textbf{P}_\ell^\bot \left(\bar{\textbf{e}}(t) - \textbf{e}_\ell(t)\right),  \\ \label{e_recur2} &&  \bar{\textbf{e}}(t+1) = \frac{\eta}{M} \sum_{\ell=1}^M \textbf{e}_\ell(t+1) + (1 - \eta ) \bar{\textbf{e}}(t).
\end{eqnarray}

It follows from (\ref{appx_L1_2}) that $y_\ell = \textbf{A}_\ell \textbf{x}^* + \hat{w}_\ell$, while every $\textbf{x}_{\ell}(t)$ computed in Line 5 or Line 10 of Algorithm \ref{algorithm2} should be a solution of $y_\ell = \textbf{A}_\ell \textbf{x}$, i.e., \begin{eqnarray}y_\ell = \textbf{A}_\ell \textbf{x}_{\ell}(t).\end{eqnarray}
This can be verified using the fact that, as long as $\textbf{A}_\ell \textbf{x}_\ell(0) = y_\ell$, $\textbf{A}_\ell \textbf{x}_\ell(t+1) = \textbf{A}_\ell \textbf{x}_\ell(t) + \gamma \textbf{A}_\ell \textbf{P}_\ell^\bot \left(\overline{\textbf{x}}(t) - \textbf{x}_\ell(t) \right) = \textbf{A}_\ell \textbf{x}_\ell(t) = y_\ell$ holds true for all $t \geq 0$, according to Lemma 1. Therefore,
\begin{eqnarray} \label{LMA_A1_post} \textbf{P}_\ell^\bot \textbf{e}_\ell(t) &=& \textbf{e}_\ell(t) - \textbf{A}_\ell^H \left( \textbf{A}_\ell \textbf{A}_\ell^H \right)^{-1}  \left( \textbf{A}_\ell \textbf{x}_{\ell}(t) - \textbf{A}_\ell \textbf{x}^* \right) \nonumber \\ &=& \textbf{e}_\ell(t) - \textbf{A}_\ell^H \left( \textbf{A}_\ell \textbf{A}_\ell^H \right)^{-1} \tilde{w}_\ell, \end{eqnarray}
which allows us to rewrite (\ref{e_recur1}) as
\begin{eqnarray}  \label{e_recur1a}
\textbf{e}_\ell(t+1) = (1 - \gamma)\textbf{e}_\ell(t) + \gamma \textbf{P}_\ell^\bot \bar{\textbf{e}}(t) + \gamma \textbf{A}_\ell^H \left( \textbf{A}_\ell \textbf{A}_\ell^H \right)^{-1} \tilde{w}_\ell,
\end{eqnarray}

From (\ref{e_recur2}) and (\ref{e_recur1a}), we can develop a state-space equation to describe the key recursions of the APC algorithm as follows:
\begin{eqnarray} \label{dt_equ}
\textbf{d}(t+1) = \textbf{G} \textbf{d}(t) + \tilde{\textbf{w}}_d,
\end{eqnarray}
where
\begin{eqnarray} \label{app_def_d0}
&& \textbf{d}(t) = \left[
      \begin{array}{c}
        \textbf{e}_1(t) \\
        \vdots\\
        \textbf{e}_M(t) \\
        \bar{\textbf{e}}(t) \\
      \end{array}
    \right], \quad  \tilde{\textbf{w}}_d = \gamma \left[
      \begin{array}{c}
         \textbf{A}_1^H \left( \textbf{A}_1 \textbf{A}_1^H \right)^{-1} \tilde{w}_1 \\
        \vdots\\
         \textbf{A}_M^H \left( \textbf{A}_M \textbf{A}_M^H \right)^{-1} \tilde{w}_M \\
        \textbf{0}_s \\
      \end{array}
    \right],  \quad\quad \label{dt_equ1_0} \\  &&
\textbf{G} = \left[
      \begin{array}{cc}
        (1-\gamma) \textbf{I}_{M s} & \gamma \left[
                                          \begin{array}{c}
                                            \textbf{P}_{1}^\bot \\
                                            \vdots \\
                                            \textbf{P}_{M}^\bot \\
                                          \end{array}
                                        \right]
         \\
        \frac{\eta (1-\gamma)}{M} \left[\begin{array}{ccc}
                                    \textbf{I}_{s} & \cdots & \textbf{I}_{s}
                                  \end{array} \right]
         & \textbf{B} \\
      \end{array}
    \right], \label{dt_equ1}
    \end{eqnarray}
with $\textbf{B} = \frac{\eta \gamma}{M} \sum_{\ell = 1}^M  \textbf{P}_{\ell}^\bot + (1 - \eta) \textbf{I}_{s} =  - \eta \gamma \textbf{X} + (1 - \eta + \eta \gamma) \textbf{I}_{s} $.

\emph{Lemma 2}: $\textbf{G}$ have $(M + 1) s$ eigenvalues, among which there are  $(M - 1) s$ eigenvalues that are equal to $1-\gamma$, and $2 s$ eigenvalues $\xi_{1, \pm}, \cdots, \xi_{s, \pm}$, where $\xi_{i, \pm}$ ($i=1, \cdots, s$) are the solutions of the quadratic equation \begin{eqnarray} \label{lambda_quad}\xi^2 + (- \eta \gamma (1-\theta_i) + \gamma + \eta - 2)\xi + (\gamma-1)(\eta-1) = 0,\end{eqnarray} such that
\begin{eqnarray} \label{lambda_pm_def} \xi_{i, \pm} = \frac{\theta_{max} + \theta_{min} - 2\theta_i}{\left(\sqrt{\theta_{max}}+ \sqrt{\theta_{min}} \right)^2}  \pm  \frac{2 \sqrt{(\theta_i - \theta_{max})(\theta_i - \theta_{min})}}{\left(\sqrt{\theta_{max}}+ \sqrt{\theta_{min}} \right)^2}.
\end{eqnarray}
If $\theta_{min} < \theta_i <  \theta_{max}$, then $\left(\theta_i - \theta_{max} \right) \left(\theta_i - \theta_{min}\right) < 0$ and thus $\xi_{i, \pm}$ are complex-valued.

\emph{Lemma 3}: For $\textbf{G}$, the spectral radius $\rho(\textbf{G}) = \alpha <1$, so that $\lim_{t \rightarrow \infty} \textbf{G}^t = \textbf{0}_{(M+1) s}$, where $\alpha$ is defined in (\ref{alpha_def1}), and $t$ is a positive integer. Besides, the Neumann series $\sum_{l=0}^{\infty} \textbf{G}^{l}$ converges, i.e., $\sum_{l=0}^{\infty} \textbf{G}^{l} = (\textbf{I}-\textbf{G})^{-1}$.

Because the system (\ref{dt_equ}) is in the form of a discrete-time state-space equation, its closed-form solution can be directly obtained by applying the linear system theory \cite{Chen_1999}.

\textbf{\emph{Theorem 1 \cite[(4.20)]{Chen_1999}}}: The solution to the system (\ref{dt_equ}) can be written as
\begin{eqnarray} \label{dt_solu}
\textbf{d}(t) = \textbf{G}^t \textbf{d}(0) + \left(\sum_{l=0}^{t-1} \textbf{G}^{l}\right) \tilde{\textbf{w}}_d,\label{d_dai}
\end{eqnarray}
where $\textbf{G}^t \textbf{d}(0)$ and $\left(\sum_{l=0}^{t-1} \textbf{G}^{l}\right) \tilde{\textbf{w}}_d $ are the zero-input and zero-state responses, respectively.

Use Lemma 3 to show that
\begin{eqnarray}&&\sum_{l=0}^{t-1} \textbf{G}^{l} = \sum_{l=0}^{\infty} \textbf{G}^{l}- \textbf{G}^{t} \left(\sum_{l=0}^{\infty} \textbf{G}^{l} \right) = \left(\textbf{I} - \textbf{G}^{t}\right)(\textbf{I}-\textbf{G})^{-1}. \quad\quad \end{eqnarray}
Substituting this result into (\ref{dt_solu}) produces
\begin{eqnarray}\label{d_dai_c}
\textbf{d}(t) = \textbf{G}^t \textbf{d}(0) + \left(\textbf{I} - \textbf{G}^{t}\right) \left(\textbf{I}-\textbf{G}\right)^{-1} \tilde{\textbf{w}}_d,
\end{eqnarray}
which suggests that
\begin{eqnarray} \label{d_dai_infty}
\textbf{d}(\infty) := \lim_{t \rightarrow \infty} \textbf{d}(t) = \left(\textbf{I}-\textbf{G}\right)^{-1} \tilde{\textbf{w}}_d.
\end{eqnarray}
 Observe that the behavior of $\textbf{d}(t)$ relies heavily on $\textbf{G}$ and $\textbf{G}^t$, so we will apply the Jordan canonical form theorem to the coming analysis.

\emph{Theorem (Jordan Canonical Form \cite[Theorem 3.1.11]{Horn_1985})}: There exists a nonsingular matrix $\textbf{S} \in \mathbb{C}^{(M+1) s \times (M+1) s}$, and there are positive integers $q$, $n_1,\cdots, n_q$ with $n_1 + \cdots +n_q = (M+1) s$, and scalars $\xi_1, \cdots, \xi_q \in \{1-\gamma, \xi_{1, \pm}, \xi_{2, \pm}, \cdots, \xi_{s, \pm}\}$ such that \begin{eqnarray} \label{thm_a_1}\textbf{G} = \textbf{S}^{-1} \textbf{J} \textbf{S},\end{eqnarray} where $\textbf{J} = \textbf{J}_{n_1} (\xi_1) \bigoplus \cdots \bigoplus \textbf{J}_{n_q} (\xi_q)$ is a Jordan matrix, and $\textbf{J}_{n_l} (\xi_l)$, $l=1,\cdots,q$, are Jordan blocks\footnote{The Jordan block $\textbf{J}_{n_l} (\xi_l)$ is an $n_l$-by-$n_l$ upper triangular matrix in which $\xi_l$ appears $n_l$ times on the main diagonal; if $n_l > 1$, there are $n_l-1$ elements $1$ in the superdiagonal; all other elements are $0$ \cite{Horn_1985}.}.

This theorem has an important consequence. Precisely, we have the following corollary.

\emph{Corollary 1}: Let $\textbf{G}$ be given. Then $\textbf{G}^t = \textbf{S}^{-1} \textbf{J}^t \textbf{S}$.

Next, we investigate the properties of $\textbf{J}^t$ by establishing two lemmas (i.e., Lemmas 4 and 6 given below) that identify the features of Jordan blocks with different eigenvalues of $\textbf{G}$.

\emph{Lemma 4 (Jordan Blocks with Eigenvalue $1-\gamma$ of $\textbf{G}$)}: The number of Jordan blocks of $\textbf{G}$ corresponding to eigenvalue $1-\gamma$ is $(M-1) s$, such that every Jordan block is $1$-by-$1$.

The next result aims to provide insights into Jordan blocks with eigenvalues $\xi_{i, \pm}$, $i=1, \cdots, s$. Before stating this result, we will first give an explicit formula for the eigenvectors associated with these eigenvalues.

\emph{Lemma 5}: Let $\bm{\Lambda}=\textrm{diag}(\theta_1, \cdots, \theta_s)$, where $\theta_1, \cdots, \theta_s$ are the eigenvalues of $\textbf{X}$, see (\ref{def_X_0425}) and (\ref{def_theta_0425}). Then
\begin{description}
  \item[\emph{(a)}] $\textbf{X}$ has $s$ orthonormal eigenvectors, denoted by $\textbf{v}_1, \cdots, \textbf{v}_s$, where $\textbf{v}_i$ is the eigenvector associated with eigenvalue $\theta_i$ of $\textbf{X}$, $i=1,\cdots, s$.
  \item[\emph{(b)}] Every scalar $\xi \in \left \{\xi_{1, \pm},\cdots,\xi_{s, \pm} \right\}$ (such that $\xi$ is a solution to (\ref{lambda_quad})) and
\begin{eqnarray}
\textbf{v}_G(\xi) = \left[\begin{array}{c}
                                            - \frac{\gamma}{1-\gamma - \xi} \textbf{P}_{1}^\bot \textbf{v}_i\\
                                            - \frac{\gamma}{1-\gamma - \xi} \textbf{P}_{2}^\bot \textbf{v}_i\\
                                            \vdots \\
                                            \textbf{v}_i \\
                                          \end{array}\right],
\end{eqnarray}
form an eigenvalue-eigenvector pair for $\textbf{G}$.
\end{description}

\emph{Lemma 6 (Jordan Blocks with Eigenvalues $\xi_{i, \pm}$, $i=1, \cdots, s$, of $\textbf{G}$)}: The number of Jordan blocks of $\textbf{G}$ corresponding to every eigenvalue $\xi \in \{\xi_{1, \pm},\cdots, \xi_{s, \pm}\}$ is at most $2$, where each Jordan block is either $1$-by-$1$ or $2$-by-$2$.

The preceding results (including Lemmas 4 and 6) motivate the property of $\textbf{J}^t$, as shown by the following lemma.

\emph{Lemma 7}: For the Jordan matrix $\textbf{J}$ in (\ref{thm_a_1}), we have \begin{eqnarray} \label{thmA2_1} &&\textbf{J}^t = \textbf{J}_{n_1}^t (\xi_1) \bigoplus \cdots \bigoplus \textbf{J}_{n_q}^t (\xi_q), \end{eqnarray} where
\begin{eqnarray}
\textbf{J}_{n}^t(\xi) = \left[
                     \begin{array}{cccc}
                       \xi^t & \binom{t}{1}
                        \xi^{t-1} & \cdots & \binom{t}{n-1} \xi^{t-n+1}\\
                       0 & \xi^t & \cdots & \binom{t}{n-2} \xi^{t-n+2} \\
                       \vdots & \vdots & \ddots & \vdots \\
                       0 & 0 & \cdots & \xi^t \\
                     \end{array}
                   \right],
\end{eqnarray}
and $n_l \in \{1, 2\}$ for all $l=1,\cdots,q$.

After obtaining Corollary 1 and Lemma 7, the mechanism governing the convergence of $\textbf{G}^t$ (in (\ref{d_dai_c})) to zero is better understood. Now we are in a position to give a formulation that characterizes the behavior of $\textbf{d}(t)$.

\emph{Theorem 2}: Consider the solution $\textbf{d}(t)$ of the state-space equation (\ref{dt_equ}). We have \begin{eqnarray} \label{d_A1} \textbf{d}(t) = \textbf{d}(\infty) + \textbf{G}^t \left(\textbf{d}(0)  - \left(\textbf{I}_{(M+1) s}-\textbf{G}\right)^{-1} \tilde{\textbf{w}}_d \right),\end{eqnarray} since $\textbf{d}(\infty) = \left(\textbf{I}_{(M+1) s}-\textbf{G}\right)^{-1} \tilde{\textbf{w}}_d$, such that
\begin{eqnarray} \label{epsilon_eq}
  \textbf{d}(t) = \left(\textbf{I}_{(M+1) s}-\textbf{G}\right)^{-1} \tilde{\textbf{w}}_d + \left(\textbf{I}_{(M+1) s}-\textbf{G}\right)^{-1} \bm{\epsilon}, \quad\quad
\end{eqnarray}
where \begin{eqnarray} \label{d_norm_infty_ine} \bm{\epsilon}=\textbf{G}^t \left(\left(\textbf{I}_{(M+1) s}-\textbf{G}\right) \textbf{d}(0)  - \tilde{\textbf{w}}_d \right).\end{eqnarray}Moreover, $\| \bm{\epsilon} \|_2 = O \left( \alpha^{t} \right)$.

\emph{Theorem 3}: Consider the problem of solving a large-scale system (\ref{appx_L1_1}) of linear equations, we have\begin{eqnarray} \label{thm_A2} \bar{\textbf{x}}(t) - \textbf{x}^* = \frac{1}{(1+\eta) M} \textbf{X}^{-1} \bm{A}^H \bm{\Xi} \tilde{\textbf{w}} + \frac{1}{\gamma(1+\eta)} \textbf{X}^{-1} \bm{\bar{\epsilon}}, \end{eqnarray}
where \begin{eqnarray} \label{epsilon_bar_eq} \bm{\bar{\epsilon}}= \left[\begin{array}{cccc}\frac{1}{M}\textbf{I}_{s} & \frac{1}{M}\textbf{I}_{s} & \cdots & \textbf{I}_{s}\end{array} \right] \epsilon.\end{eqnarray} 

Theorem 3 characterizes the error of solution attained by the APC algorithm when the system (\ref{appx_L1_1}) is faced with unknown noise $\tilde{\textbf{w}}$ as in (\ref{appx_L1_2}). It is easy to see that Theorem 3 is a generalization of Theorem 1 derived in \cite{Azizan_2019}, because by letting $\tilde{\textbf{w}}$ be the all-zero vector in Theorem 3, our obtained result can be reduced to Theorem 1 of \cite{Azizan_2019}.

\section{Numerical Results}

In this section, we carry out a serise of simulations to evaluate the performance of the APC algorithm in the noise scenarios with the aim to verify Theorem 3. By changing iteration times $T$, we focus on the mean square error (MSE) performance which defined as:
\begin{eqnarray}\label{appx_L1_}
\rm{MSE} = \mathbb{E} {\left \{{\left\| {\mathbf{x}^*}-\tilde{\mathbf{x}} \right\|_{\rm{2}}^2} \right \}}.
\end{eqnarray}
We mainly consider the influence of three factors: the size of matrix \textbf{A} (i.e., the number of $M$), the condition number of matrix \textbf{X} (i.e., $\kappa \left(\textbf{X} \right)$), and the noise power $P_n$.

\begin{figure}[htbp]
\centerline{\includegraphics[width = 0.45\textwidth]{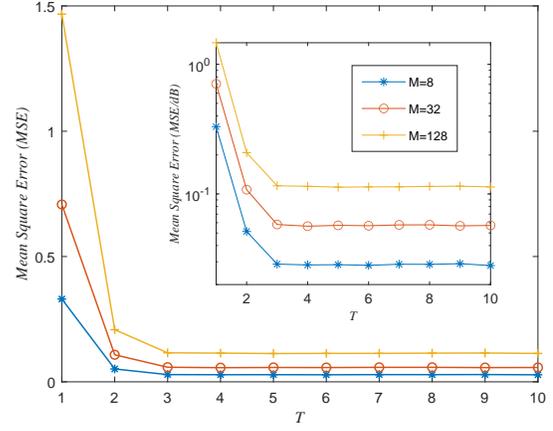}}
\caption{MSE performance of APC algorithm as a function of $T$ with different $M$.}
\label{fig1}
\end{figure}

Fig.\ref{fig1} shows the MSE performance of APC algorithm with different size of matrix \textbf{A}. We set $\kappa \left(\textbf{X} \right)$ = 1.6, and the noise power $P_n=0.0001$. When $M=\left \{8,32,128 \right \}$, the simulation results are shown in Fig.\ref{fig1}. We can find that no matter what $M$ is equal to, APC can approach convergence after $T=4$. With the increase of the $M$, the MSE performance decreases.

\begin{figure}[htbp]
\centerline{\includegraphics[width = 0.45\textwidth]{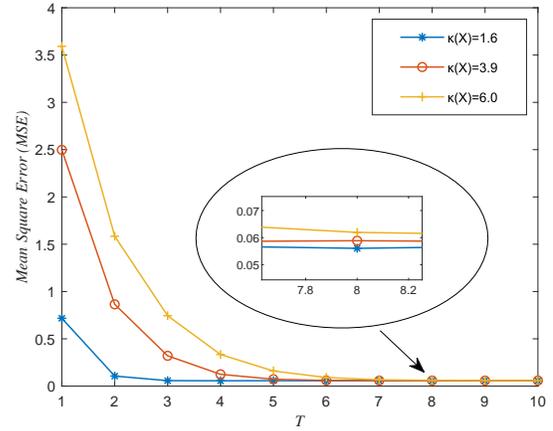}}
\caption{MSE performance of APC algorithm as a function of $T$ with different $\kappa \left(\textbf{X} \right)$.}
\label{fig2}
\end{figure}

Fig.\ref{fig2} is plotted when $M=32$ and $P_n=0.0001$. It shows the influence of $\kappa \left(\textbf{X} \right)$ on MSE performance. By comparing the curves, we can observe that if we keep $M$ and $P_n$ the same, the degree of convergence of the APC algorithm varies only slightly. But there is large difference in convergence speed. When the condition number is larger, the convergence speed is lower. Specifically, when $\kappa \left(\textbf{X} \right)=1.56$, algorithm converges after 3 iterations; but when $\kappa \left(\textbf{X} \right)$ increases to 6.0, algorithm needs 9 iterations to get convergence.


\begin{figure}[htbp]
\centerline{\includegraphics[width = 0.45\textwidth]{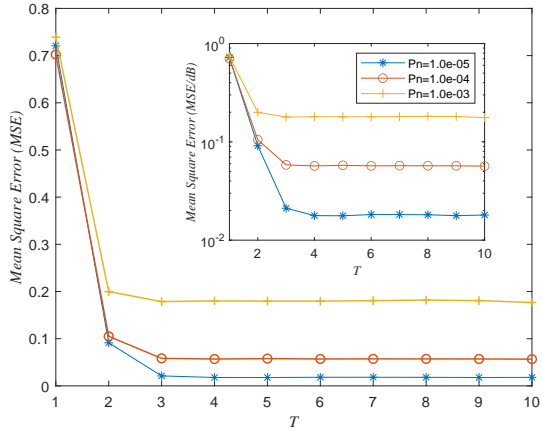}}
\caption{MSE performance of APC algorithm as a function of $T$ with different $P_n$.}
\label{fig3}
\end{figure}

Fig.\ref{fig3} illustrates the impact of the noise power $P_n$ on the MSE performance for APC algorithm. We gradually increase $P_n$ from 0.00001 to 0.001. To be expected, the increase of $P_n$ won’t make influence on convergence speed, but MSE performance will getting worse.

As a summary, as the increase of $T$, $\frac{1}{\gamma(1+\eta)} \textbf{X}^{-1} \bm{\bar{\epsilon}}$ in \eqref{thm_A2} tends to 0, the error is only effected by $\frac{1}{(1+\eta) M} \textbf{X}^{-1} \bm{A}^H \bm{\Xi} \tilde{\textbf{w}}$ in \eqref{thm_A2}. In other words, the error of APC algorithm mainly determined by the noise level after enough iterations. This fits well with Theorem 3.


\section{Conclusion}

In this paper, we generalize the analysis of the APC algorithm, and clarify the error performance of the algorithm in presence of noise for solving linear systems. The generalization should be necessary to refine the theoretical framework of the APC algorithm. Also, we provide closed-form expressions to two important parameters of APC.

\section*{Acknowledgment}

This work is sponsored in part by the National Natural Science Foundation of China (grant no. 61801048, 61971058, 62071063) and Beijing Natural Science Foundation (grant no. L202014, L192002).

\section*{Appendix: Proofs of Main Results}

\emph{Proof of Lemma 3}: A computation reveals that \begin{eqnarray}|\xi_{i, \pm}|= \sqrt{(\gamma-1)(\eta-1)} = \alpha,\end{eqnarray} for all $i=1, \cdots, s$. Then, it's a simple consequence of Lemma A.2 that $\rho(\textbf{G}) = \alpha <1$ \cite{Azizan_2019}, i.e., the largest magnitude eigenvalue of $\textbf{G}$ is less than $1$. This implies that $\lim_{t \rightarrow \infty} \textbf{G}^t = \textbf{O}_{(M+1) s}$ and $\sum_{l=0}^{\infty} \textbf{G}^{l} = (\textbf{I}-\textbf{G})^{-1}$ \cite{Meyer_2000}. $\quad\qquad\qquad\qquad\qquad\qquad \quad \blacksquare$

\emph{Proof of Lemma 4}: All we need to do is to show that the geometric multiplicity of eigenvalue $1-\gamma$ is $(M-1) s$ \cite{Horn_1985,Meyer_2000}. We begin by noting that
\begin{eqnarray}
\textbf{G} - (1-\gamma) \textbf{I} &=& \left[
      \begin{array}{cc}
        \textbf{O}_{M s} & \gamma \left[
                                          \begin{array}{c}
                                            \textbf{P}_{1}^\bot \\
                                            \vdots \\
                                            \textbf{P}_{M}^\bot \\
                                          \end{array}
                                        \right]
         \\
        \frac{\eta (1-\gamma)}{M} \left[\begin{array}{ccc}
                                    \textbf{I}_{s} & \cdots & \textbf{I}_{s}
                                  \end{array} \right]
         & \textbf{B} - (1-\gamma) \textbf{I}_{s} \\
      \end{array}
    \right]. \nonumber
\end{eqnarray}
Applying elementary row and column operations \cite{Horn_1985}, we can transform $\textbf{G} - (1-\gamma) \textbf{I}$ into a simple form
\begin{eqnarray}
 \textbf{G}_T = \left[
      \begin{array}{cc}
        \textbf{O}_{M s} & \gamma \left[
                                          \begin{array}{c}
                                            \textbf{P}_{1}^\bot \\
                                            \vdots \\
                                            \textbf{P}_{M}^\bot \\
                                          \end{array}
                                        \right]
         \\
        \frac{\eta (1-\gamma)}{M} \left[\begin{array}{cccc}
                                    \textbf{O}_{s} & \cdots & \textbf{O}_{s} & \textbf{I}_{s}
                                  \end{array} \right]
         & \textbf{O}_{s} \\
      \end{array}
    \right]. \nonumber
\end{eqnarray}
Since elementary operations do not change the rank of a matrix \cite{Horn_1985}, it follows that\begin{eqnarray} \label{geom_mul}\textmd{rank} \left(\textbf{G} - (1-\gamma) \textbf{I} \right) = \textmd{rank} \left(\textbf{G}_T\right)  \leq 2 s.\end{eqnarray} The geometric multiplicity of eigenvalue $1-\gamma$ of $\textbf{G}$ is equal to $(M+1) s - \textmd{rank} \left(\textbf{G} - (1-\gamma) \textbf{I} \right)$ that is not less than $(M-1) s$ according to (\ref{geom_mul}). Furthermore, the geometric multiplicity should not be larger than algebraic multiplicity, i.e., $(M-1) s$, for $1-\gamma$, and thus can only be $(M-1) s$ \cite[Theorem 1.4.10]{Horn_1985}. This implies that the number of Jordan blocks of $\textbf{G}$ corresponding to $1-\gamma$, is $(M-1) s$ \cite{Horn_1985}. Finally, because the geometric and algebraic multiplicities of $1-\gamma$ are equal, every Jordan block corresponding to $1-\gamma$ is $1$-by-$1$ \cite{Horn_1985}. $\qquad\qquad \blacksquare$

\emph{Proof of Lemma 5}: First note that $\textbf{X}$ is Hermitian and $\textbf{X}$ is unitarily diagonalizable \cite[Theorem 2.5.6]{Horn_1985}. Then, applying \cite[Theorem 2.5.3]{Horn_1985} yields the assertion (a).

To prove the assertion (b), let us verify whether $\textbf{v}_G(\xi)$ satisfies the eigenvalue-eigenvector equation $(\textbf{G} - \xi \textbf{I}) \textbf{v}_G(\xi)  = \textbf{0}_{(M+1) s}$, where
\begin{eqnarray}
&& \textbf{G} - \xi \textbf{I}_{(M+1) s} = \left[
      \begin{array}{cc}
        (1-\gamma - \xi) \textbf{I}_{M s} & \gamma \left[
                                          \begin{array}{c}
                                            \textbf{P}_{1}^\bot \\
                                            \vdots \\
                                            \textbf{P}_{M}^\bot \\
                                          \end{array}
                                        \right]
         \\
        \frac{\eta (1-\gamma)}{M} \left[\begin{array}{ccc}
                                    \textbf{I}_{s} & \cdots & \textbf{I}_{s}
                                  \end{array} \right]
         & \textbf{B} - \xi \textbf{I}_{s} \\
      \end{array}
    \right]. \nonumber
\end{eqnarray}

On the one hand, it is easy to check that
\begin{eqnarray} \label{v_G_s1}
\left[
      \begin{array}{cc}
        (1-\gamma - \xi) \textbf{I}_{M L} & \gamma \left[
                                          \begin{array}{c}
                                            \textbf{P}_{1}^\bot \\
                                            \vdots \\
                                            \textbf{P}_{M}^\bot \\
                                          \end{array}
                                        \right]
         \\
      \end{array}
    \right] \textbf{v}_G(\xi) = \textbf{0}_{M s}.
\end{eqnarray}
On the other hand, since $\textbf{X} \textbf{v}_i = \theta_i \textbf{v}_i$ and $\frac{\sum_{\ell = 1}^M  \textbf{P}_{\ell}^\bot}{M} = \textbf{I}_{s} - \textbf{X}$, we have $\frac{\sum_{\ell = 1}^M  \textbf{P}_{\ell}^\bot}{M} \textbf{v}_i = (1 - \theta_i) \textbf{v}_i$ and $\left(\textbf{B} - \xi \textbf{I}_{s}\right)\textbf{v}_i = (- \eta \gamma \theta_i +1 - \eta + \eta \gamma - \xi) \textbf{v}_i$. As (\ref{lambda_quad}) ensures that $-\frac{\eta \gamma (1-\gamma)}{1-\gamma - \xi} (1 - \theta) - \eta \gamma \theta + 1 - \eta + \eta \gamma - \xi = 0$ if $\xi \neq 1-\gamma$, it yields
\begin{eqnarray} \label{v_G_s2}
&&\left[
      \begin{array}{cc}
        \frac{\eta (1-\gamma)}{M} \left[\begin{array}{ccc}
                                    \textbf{I}_{s} & \cdots & \textbf{I}_{s}
                                  \end{array} \right]
         & \textbf{B} - \xi \textbf{I}_{s} \\
      \end{array}
    \right] \textbf{v}_G(\xi) \nonumber \\ && \quad = \left[- \frac{\eta \gamma (1-\gamma)}{(1-\gamma - \xi)} \frac{\sum_{\ell = 1}^M  \textbf{P}_{\ell}^\bot}{M} + \left(\textbf{B} - \xi \textbf{I}_{s}\right) \right] \textbf{v}_{i} = \textbf{0}_{s}. \quad\quad
\end{eqnarray}
Finally, (\ref{v_G_s1}) and (\ref{v_G_s2}) together imply that $(\textbf{G} - \xi \textbf{I}) \textbf{v}_G(\xi)  = \textbf{0}_{(M+1) s}$ which completes the proof of the assertion (b). $\quad\quad\,\,\,\, \blacksquare$

\emph{Proof of Lemma 6}: It follows from (\ref{lambda_pm_def}) that $\xi_{1, +} = \xi_{1, -}$ and $\xi_{s, +} = \xi_{s, -}$ since $\theta_s = \theta_{min}$ and $\theta_1 = \theta_{max}$, while $\xi_{i, +} \neq \xi_{i, -}$ if $\theta_{min} < \theta_i <  \theta_{max}$. One of the consequences of Lemma 5 is that the number of linearly dependent eigenvectors associated with every eigenvalue $\xi \in \{\xi_{1, \pm},\cdots,\xi_{s, \pm}\}$ is not larger than $2$. Therefore, the geometric multiplicity of $\xi$ is $1$, and the Jordan block of $\textbf{G}$ with eigenvalue $\xi$ is $2$-by-$2$, when $\xi = \xi_{1, +} = \xi_{1, -}$ or $\xi = \xi_{s, +} = \xi_{s, -}$. If $\xi = \xi_{i, +}$ or $\xi = \xi_{i, -}$ with $\xi_{i, +} \neq \xi_{i, -}$, then the geometric multiplicity of $\xi$ is $2$ such that the Jordan block with eigenvalue $\xi$ is $1$-by-$1$. $\,\,\,\,\qquad\qquad\qquad\qquad\quad\quad\quad\quad\,\, \blacksquare$

\emph{Proof of Lemma 7}: It can easily be verified by using the property of direct sum together with \cite[(7.10.7)]{Meyer_2000}. $\qquad \qquad \blacksquare$

\emph{Proof of Theorem 2}: Combining (\ref{d_dai_c}) and (\ref{d_dai_infty}) yields (\ref{d_A1}), and then multiplying on both sides of (\ref{d_A1}) by $\textbf{I}_{(M+1) s}-\textbf{G}$ produces (\ref{d_norm_infty_ine}). By the Rayleigh quotient theorem \cite{Meyer_2000} and Corollary 1, we obtain \begin{eqnarray} \label{prf_thmA1} \left \| \bm{\epsilon} \right\|_2 \leq \left \| \textbf{J}^t \right \|_2 \left \| \left(\textbf{I}_{(M+1) s}-\textbf{G}\right) \textbf{d}(0)  - \tilde{\textbf{w}}_d  \right \|_2,\end{eqnarray}
where $\left \| \textbf{J}^t \right \|_2 = O \left( \alpha^{t} \right)$ \cite{Chen_1999}.

The remaining part of the proof is to derive an upper bound on $\left \| \left(\textbf{I}_{(M+1) s}-\textbf{G}\right) \textbf{d}(0)  - \tilde{\textbf{w}}_d  \right \|_2$. First, it is clear that
\begin{eqnarray} \label{app_d__0}
 \textbf{d}(0) = \left[
      \begin{array}{c}
        -\textbf{P}_1^\bot \textbf{x}^* + \textbf{A}_1^H \left( \textbf{A}_1 \textbf{A}_1^H \right)^{-1} \tilde{w}_1 \\
        \vdots\\
        -\textbf{P}_M^\bot \textbf{x}^* + \textbf{A}_M^H \left( \textbf{A}_M \textbf{A}_M^H \right)^{-1} \tilde{w}_M \\
        - \frac{1}{M} \sum_{\ell=1}^M \textbf{P}_\ell^\bot \textbf{x}^* + \frac{1}{M} \sum_{\ell=1}^M \textbf{A}_\ell^H \left( \textbf{A}_\ell \textbf{A}_\ell^H \right)^{-1} \tilde{w}_\ell \\
      \end{array}
    \right],
    \end{eqnarray}
according to the initialization of the APC algorithm as in Line 5 of Algorithm \ref{algorithm2} together with (\ref{app_def_d0}), and thus
\begin{eqnarray}
\left(\textbf{I}_{(M+1) s}-\textbf{G}\right) \textbf{d}(0)  - \tilde{\textbf{w}}_d = \left[
      \begin{array}{c}
        - \gamma \textbf{P}_1^\bot  \left(\textbf{x}^* + \bar{\textbf{e}}(0) \right)\\
        \vdots\\
         - \gamma \textbf{P}_M^\bot  \left(\textbf{x}^* + \bar{\textbf{e}}(0) \right) \\
        - \eta \gamma \textbf{X} \bar{\textbf{e}}(0) \\
      \end{array}
    \right]. \nonumber
\end{eqnarray}
 This yields
\begin{eqnarray}
&& \left \| \left(\textbf{I}_{(M+1) s}-\textbf{G}\right) \textbf{d}(0)  - \tilde{\textbf{w}}_d \right \|_2^2 \leq 2 \gamma^2 M  \left(1 - \theta_{min} \right)  \left \| \textbf{x}^* \right \|_2^2 \nonumber \\ && \quad \quad + \left(2 \gamma^2 M \left(1 - \theta_{min} \right) + \eta^2 \gamma^2 \theta_{max}^2  \right) \left \| \bar{\textbf{e}}(0) \right \|_2^2. \nonumber
\end{eqnarray}

 Then, since $\left \| \bar{\textbf{e}}(0) \right \|_2^2 \leq \frac{\sum_{\ell=1}^M \left \| \textbf{e}_\ell (0) \right \|_2^2}{M}$, it follows from Lemma 1 and (\ref{app_d__0}) that
\begin{eqnarray}
 \left \| \bar{\textbf{e}}(0) \right \|_2^2 &\leq&
 \frac{M \left(\textbf{x}^* \right)^H \left(\textbf{I} - \textbf{X}\right) \textbf{x}^* + \sum_{\ell = 1}^M \left( \textbf{A}_\ell \textbf{A}_\ell^H \right)^{-2} \tilde{w}_\ell}{M} \nonumber \\  & \leq &  \left(1 - \theta_{min} \right) \left \|\textbf{x}^* \right\|^2_2 + \frac{\sum_{\ell = 1}^M \left( \textbf{A}_\ell \textbf{A}_\ell^H \right)^{-2} \tilde{w}_\ell}{M},\nonumber   \end{eqnarray}
and
\begin{eqnarray}
&& \left \| \left(\textbf{I}_{(M+1) s}-\textbf{G}\right) \textbf{d}(0)  - \tilde{\textbf{w}}_d \right \|_2 \leq \left(2 \gamma^2 M \left(1 - \theta_{min} \right) + \eta^2 \gamma^2 \theta_{max}^2  \right)  \nonumber \\  && \quad \quad \times \left(\left(2 - \theta_{min} \right) \left \|\textbf{x}^* \right\|^2_2 + \frac{\sum_{\ell = 1}^M \left( \textbf{A}_\ell \textbf{A}_\ell^H \right)^{-2} \tilde{w}_\ell}{M} \right). \nonumber
\end{eqnarray}
Substituting this inequality into (\ref{prf_thmA1}) gives $\| \bm{\epsilon} \|_2 = O \left( \alpha^{t} \right)$. $\quad \,\,\blacksquare$

\emph{Proof of Theorem 3}: Observe that \begin{eqnarray} && \left[\begin{array}{cccc} \frac{1}{M}\textbf{I}_{s} & \frac{1}{M}\textbf{I}_{s} & \cdots & \textbf{I}_{s}\end{array} \right] \textbf{d}(t) = 2 \bar{\textbf{e}}(t), \nonumber \\ && \left[\begin{array}{cccc}\frac{1}{M}\textbf{I}_{s} & \frac{1}{M}\textbf{I}_{s} & \cdots & \textbf{I}_{s}\end{array} \right] \textbf{G} \textbf{d}(t) = \left(2\textbf{I}_{s} - \gamma(1+\eta)\textbf{X} \right) \bar{\textbf{e}}(t). \nonumber \end{eqnarray} A calculation also shows that $\left[\begin{array}{cccc}\frac{1}{M}\textbf{I}_{s} & \frac{1}{M}\textbf{I}_{s} & \cdots & \textbf{I}_{s}\end{array} \right] \tilde{\textbf{w}}_d = \frac{\gamma}{M} \bm{A}^H \bm{\Xi} \tilde{\textbf{w}}$. Combining these results, we get
\begin{eqnarray} \quad \gamma(1+\eta)\textbf{X} \bar{\textbf{e}}(t) = \frac{\gamma}{M} \bm{A}^H \bm{\Xi} \tilde{\textbf{w}}  + \left[\begin{array}{cccc}\frac{1}{M}\textbf{I}_{s} & \frac{1}{M}\textbf{I}_{s} & \cdots & \textbf{I}_{s}\end{array} \right] \bm{\epsilon}, \nonumber \end{eqnarray}
which can be rewritten as (\ref{thm_A2}). From Theorem 1, it follows that $\| \bm{\bar{\epsilon}} \|_2 = O \left(\alpha^{t}\right)$. Finally, based on the definition in (\ref{e_bar}), Theorem 3 can be verified. $\qquad\qquad\qquad\qquad\qquad\quad\qquad\qquad\qquad \,\,\, \blacksquare$



\end{document}